\documentclass[%
 reprint,
 amsmath,amssymb,
 aps,
 prd,
 unsortedaddress,
 nofootinbib
]{revtex4-2}

\usepackage{graphicx}
\usepackage{dcolumn}
\usepackage{bm}
\usepackage{hyperref}
\usepackage{booktabs} 
\usepackage{xcolor}

\begin{document}

\preprint{APS/123-QED}

\title{The Entropic Skin: Spatial Entanglement from the QCD Confinement Boundary}

\author{Thomas B. Bahder}
 \email{tbahder@gmail.com} 
\affiliation{Quantique, LLC, 41 Southwind Drive, Belleair Bluffs, FL 33770}

\date{\today} 

\begin{abstract}
Recent investigations into High-Energy QCD have identified entanglement entropy as a crucial observable, linking parton distributions to the structure of the quantum vacuum. While momentum-space entanglement has been extensively studied in Deep Inelastic Scattering (DIS), the spatial realization of this entanglement in confined systems remains an open question. In this Letter, we demonstrate that the confining boundary of the MIT Bag Model acts as an ``Entropic Skin,'' generating maximal Spin-Position entanglement. We calculate the local reduced density matrix for the confined quark and show that the linear boundary condition, $i \gamma^\mu n_\mu \psi = \psi$, acts as an entangling gate. The surface entropy density reaches a geometric invariant of $\approx 0.918$ bits ($92\%$ of the qubit limit), independent of the bag radius. We discuss the implications of this result for Chiral Symmetry breaking and propose that this boundary entropy is the precursor to the pion cloud in effective field theories.
\end{abstract}

\maketitle

\section{Introduction}
The role of quantum information in High Energy Physics has shifted from a theoretical curiosity to a central diagnostic tool. Seminal work by Kharzeev and Levin \cite{Kharzeev2017} has established a deep connection between the parton distribution functions (PDFs) measured in Deep Inelastic Scattering and the entanglement entropy of the proton's constituent state. More recently, Sheikhi and Boroun \cite{Sheikhi2025} have extended this framework to show that the proton at high energies acts as a maximally entangled state, maximizing the entropy of the final hadronic state \cite{Blanco2024}.

Concurrently, the STAR collaboration has reported experimental evidence of vacuum entanglement via spin correlations in $\Lambda \bar{\Lambda}$ production \cite{STAR2025}, suggesting that confinement leaves a measurable imprint on quantum correlations. However, a microscopic description of \textit{how} confinement generates this entanglement in coordinate space is missing. 

Most theoretical treatments rely on infinite medium models (like NJL \cite{Nam2025}) or S-matrix formulations \cite{Miller2023}. In this work, we investigate the spatial structure of entanglement in the simplest model of confinement: the MIT Bag Model \cite{Chodos1974}. We ask: is the entanglement distributed throughout the bulk, or is it localized?

We find that the Bag Boundary ($r=R$) acts as a singular surface of information mixing. The boundary condition required to confine the color current necessarily breaks chiral symmetry and entangles the quark's spin with its position. We term this phenomenon the ``Entropic Skin'' and quantify it using the tools of Relativistic Quantum Information (RQI).

\section{Results: The Emergence of the Entropic Skin}

\subsection{The Ground State Spinor Profile}
The ground state ($1s_{1/2}$) of the MIT Bag is described by the Dirac spinor $\psi(\mathbf{r})$ \cite{DeGrand:1975cf}. Following the standard representation conventions of Landau and Lifshitz \cite{LandauQED}, the wavefunction is given by:
\begin{equation}
\psi(\mathbf{r}) = N \begin{pmatrix} j_0(E r) \chi_s \\ -i j_1(E r) (\boldsymbol{\sigma} \cdot \hat{\mathbf{r}}) \chi_s \end{pmatrix},
\label{eq:wavefunction}
\end{equation}
where $\chi_s$ is the two-component Pauli spinor, $E=x/R$ is the eigen energy defined by the bag constant $x \approx 2.04$, and $N$ is the normalization constant. The crucial physics lies in the lower component: the operator $(\boldsymbol{\sigma} \cdot \hat{\mathbf{r}})$ couples the intrinsic spin to the orbital position.

\subsection{The Local Reduced Density Matrix}
To quantify the radial dependence of spin-position entanglement, we construct the local reduced density matrix for the spin subsystem, $\rho_S(r)$. Decomposing the Dirac spinor into upper ($\phi$) and lower ($\chi$) 2-spinors, $\psi = (\phi, \chi)^T$, we trace out the angular degrees of freedom and sum over the Dirac components:
\begin{equation}
\rho_S(r) = \frac{1}{\mathcal{N}(r)} \int d\Omega \left[ \phi(\mathbf{r})\phi^\dagger(\mathbf{r}) + \chi(\mathbf{r})\chi^\dagger(\mathbf{r}) \right].
\end{equation}
Here, $\mathcal{N}(r)$ is the radial probability density, which serves as the local normalization factor to ensure $\text{Tr}(\rho_S)=1$:
\begin{equation}
\mathcal{N}(r) = \int d\Omega \, \psi^\dagger(\mathbf{r})\psi(\mathbf{r}) \propto j_0^2(xr) + j_1^2(xr).
\end{equation}
Due to the rotational symmetry of the $s$-wave ground state, the off-diagonal elements of the density matrix vanish upon integration. For an initial spin-up state ($\chi_\uparrow$), $\rho_S(r)$ becomes diagonal:
\begin{equation}
\rho_S(r) = \begin{pmatrix} P_\uparrow(r) & 0 \\ 0 & P_\downarrow(r) \end{pmatrix}.
\end{equation}
Here, $P_\downarrow(r)$ represents the probability that the spin has flipped relative to the quantization axis due to the orbital angular momentum contribution of the lower component. This mixing corresponds to the transfer of angular momentum from intrinsic spin to the orbital angular momentum of the lower component, a mechanism central to discussions of the proton spin budget \cite{Myhrer2010}.

\subsection{Analytic Derivation}
The matrix elements are determined by the relative weights of the upper ($j_0$) and lower ($j_1$) components. The angular average of $|\boldsymbol{\sigma} \cdot \hat{\mathbf{r}} \chi_\uparrow|^2$ yields a branching ratio of $1/3$ for spin-preservation and $2/3$ for spin-flip. The resulting local probabilities are:
\begin{align}
P_\uparrow(r) &= \frac{j_0^2(x r/R) + \frac{1}{3}j_1^2(x r/R)}{j_0^2(x r/R) + j_1^2(x r/R)}, \\
P_\downarrow(r) &= \frac{\frac{2}{3}j_1^2(x r/R)}{j_0^2(x r/R) + j_1^2(x r/R)}.
\end{align}

\subsection{The Two Limits}
We can now contrast the entanglement properties of the bulk and the boundary.

\textit{1. The Asymptotic Freedom Limit (Center):} In the deep interior ($r \to 0$), the quark behaves as a free particle. The lower component vanishes ($j_1 \to 0$), leading to:
\begin{equation}
\lim_{r \to 0} P_\downarrow(r) = 0.
\end{equation}
The center of the hadron is a pure product state with zero entanglement entropy ($S=0$).

\textit{2. The Confinement Limit (Surface):} At the bag surface ($r=R$), the confinement boundary condition $i \gamma^\mu n_\mu \psi = \psi$ strictly enforces $j_0(x) = j_1(x)$. Substituting this equality into our probability expressions reveals a geometric invariant:
\begin{equation}
P_\downarrow(R) = \frac{\frac{2}{3}j_0^2}{j_0^2 + j_0^2} = \frac{1}{3}.
\end{equation}
The boundary condition acts as a universal mixing gate, forcing the spin state into a fixed statistical mixture: $2/3$ Up, $1/3$ Down. The resulting von Neumann entropy is:
\begin{equation}
\mathcal{S}_{surf} = -\left( \frac{2}{3}\log_2 \frac{2}{3} + \frac{1}{3}\log_2 \frac{1}{3} \right) \approx 0.918 \text{ bits}.
\end{equation}

\begin{figure}[h]
    \centering
    \vspace{0.2in}
    \includegraphics[width=1.0\linewidth]{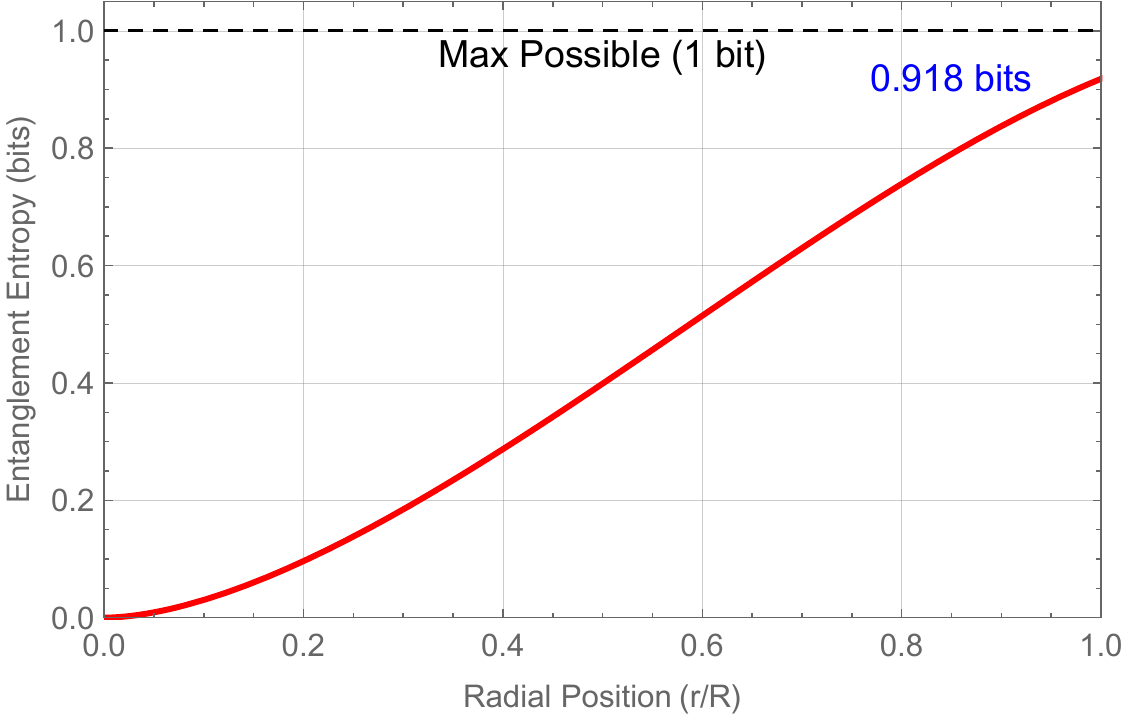}
    \caption{The emergence of the ``Entropic Skin.'' The plot shows the local von Neumann entropy of the quark spinor as a function of dimensionless radius $r/R$. The boundary condition acts as an entanglement gate, converting the purity of the interior into a mixed state at the surface. The surface value reaches $\approx 92\%$ of the theoretical maximum for a qubit.}
    \label{fig:EntanglementPlot}
\end{figure}

\begin{table*}[t]
\centering
\caption{\label{tab:skin_properties} Contrast between the asymptotically free interior and the confining surface.}
\begin{ruledtabular}
\begin{tabular}{lcc}
\textbf{Physical Property} & \textbf{Bag Interior} ($r \to 0$) & \textbf{Bag Surface} ($r \to R$) \\ \hline
Dynamical Limit & Asymptotic Freedom & Confinement \\
Wavefunction Ratio ($j_1/j_0$) & $0$ & $1$ \\
Spin-Flip Probability ($Ptheoretical proposals_\downarrow$) & $0$ & $1/3$ \\
Reduced Density Matrix & Pure State $|\chi_\uparrow\rangle$ & Mixed $\rho = \text{diag}(\frac{2}{3}, \frac{1}{3})$ \\
Local Entanglement Entropy & \textbf{0 bits} & \textbf{0.918 bits} \\
\end{tabular}
\end{ruledtabular}
\end{table*}

\section{Discussion: The Fate of the Entropic Skin}
Physically, this surface entropy can be understood as a coordinate-space realization of the vacuum entanglement observed in high-energy experiments, consistent with recent proposals of an entropic horizon derived from the QCD trace anomaly \cite{Mamo:2025hur}. By imposing a sharp boundary condition at $r=R$, the Bag Model effectively partitions the QCD vacuum, severing the correlations between the interior valence region and the exterior. The ``Entropic Skin'' of $\approx 0.918$ bits represents the irreducible uncertainty injected into the quark state by this vacuum partitioning. It acts as the geometric projection of the severed vacuum correlations.

The existence of this ``Entropic Skin'' raises a fundamental question about the restoration of Chiral Symmetry. In the static Bag Model, the chirality mixing at the boundary is hard-wired. However, in Chiral Bag Models (like the Cloudy Bag \cite{PhysRevD.22.2838,PhysRevD.24.216,Thomas1984,Owa2022}), the axial current is conserved by the emission of a pion field at the surface.

We hypothesize that the entropy we have calculated is not a static artifact, but rather the information-theoretic ``source'' for the pion cloud. We conjecture an \textbf{Entanglement Swapping} mechanism: as the chiral boundary condition is relaxed by coupling to the pion field, the Spin-Position entanglement of the quark core should decrease, being replaced by entanglement between the Quark Core and the Pion Cloud \cite{Florio2024}.

This suggests that the ``Entropic Skin'' dissolves into the bulk of the meson cloud, providing a coordinate-space realization of the vacuum entanglement observed by STAR \cite{STAR2025} and predicted by Nam \cite{Nam2025}. Future work will calculate this transfer explicitly using the Cloudy Bag Hamiltonian.

\begin{acknowledgments}
The author acknowledges the use of the large language model Gemini (Google) for assistance with Mathematica code generation and for editorial suggestions during the drafting of this manuscript.
\end{acknowledgments}

\bibliography{MITBagReferences}

\end{document}